\documentclass[aps,preprint,prl,groupaddress,amsmath,amssymb]{revtex4}

\renewcommand{\vec}[1]{{\mathbf #1}}
\newcommand{\gvec}[1]{{\boldsymbol #1}}

\usepackage{graphicx}
\usepackage{dcolumn}
\usepackage{bm}

\begin{document}

\title{Transverse confinement of waves in 3D random media\\}

\author{N. Cherroret}
\email{Nicolas.Cherroret@physik.uni-freiburg.de}
\author{S.E. Skipetrov}
\email{Sergey.Skipetrov@grenoble.cnrs.fr}
\author{B.A. van Tiggelen}
\email{Bart.Van-Tiggelen@grenoble.cnrs.fr}
\affiliation{Universit\'{e} Joseph Fourier, Laboratoire de Physique et Mod\'{e}lisation des Milieux Condens\'{e}s, CNRS UMR 5493, B.P. 166, 25 rue des Martyrs,
Maison des Magist\`{e}res, 38042 Grenoble Cedex 09, France}

\date{\today}

\begin{abstract}
We study the transmission of a tightly focused beam through a thick slab of 3D disordered medium in the Anderson localized regime. We show that the transverse profile of the transmitted beam exhibits clear signatures of Anderson localization and that its mean square width provides a direct measure of the localization length. For a short incident pulse, the width is independent of absorption.
\end{abstract}

\pacs{42.25.Dd}
\maketitle

Anderson localization is a general wave phenomenon that results from interferences of multiply scattered waves in random media and leads to exponential suppression of wave propagation beyond the localization length $\xi$ \cite{anderson58,john84,anderson85,bart99}. Recently, important advances have been made towards the observation and the theoretical description of Anderson localization of classical and, in particular, electromagnetic waves \cite{wiersma97,scheffold99,schuur99,chabanov00,chabanov03,skipetrov04,storzer06,schwarz07,zhang07}. Signatures of localization can be found in the thickness dependence of the average transmission coefficient of a random sample \cite{wiersma97}, in the shape of the coherent backscattering cone \cite{wiersma97,schuur99}, in the statistical distribution of transmitted intensity \cite{chabanov00}, or in the time-of-flight profiles of transmitted waves \cite{chabanov03,skipetrov04,storzer06,zhang07}. In a recent experiment, Schwarz \textit{et al.} \cite{schwarz07} have studied the transverse distribution of intensity of a Gaussian light beam in a medium that is translationally invariant along the direction $z$ of beam propagation, but disordered in the perpendicular $xy$ plane. As was predicted by de Raedt \textit{et al.} \cite{raedt89}, the variable $z$ plays the role of time in this case and the distribution of intensity in the $xy$ plane behaves very much as the intensity of an expanding wavepacket in a two-dimensional (2D) disordered medium. Because all waves are localized in 2D \cite{abrahams79}, the expansion of the beam is halted after one localization length $\xi$. This phenomenon was called ``transverse localization'' \cite{raedt89}.

Even though the concept of transverse localization implies 2D disorder which is translationally invariant along the direction of beam propagation, one can formally attempt the same experiment in a medium with 3D disorder. As has been shown very recently, the transverse, 2D profile of intensity of a wave transmitted through a disordered slab shows clear signatures of genuine Anderson localization in 3D \cite{hu08}. The purpose of this paper is to analyze the transverse \textit{confinement} of waves due to Anderson localization in 3D in the framework of the self-consistent theory of localization with a position-dependent diffusion coefficient \cite{tiggelen00, cherroret08}. This provides a guide for future experiments to exploit this interesting new phenomenon to access localization of various types of waves (light, sound, etc.).  We first consider a continuous monochromatic beam focused to a point at the surface of a disordered slab and show that the mean square width $\sigma^2$ of the spatially-resolved transmission coefficient  $T(\gvec{\rho})$ scales as $L \xi$ in the localized regime, when $\xi \ll L$, in contrast to the relation $\sigma^2 \propto L^2$ in the diffuse regime and at the mobility edge. This result suggests that a measurement of $\sigma^2$ gives direct access to the localization length $\xi$. However, absorption appears to produce the same effect on $\sigma^2$ as localization, i.e. in the diffuse regime we find $\sigma^2 \propto L L_a$, if the macroscopic absorption length $L_a$ is much smaller than $L$. We then replace the continuous beam by a short pulse. In the localized regime, the mean square width $\sigma^2(t)$ of the time-dependent, spatially resolved transmission coefficient $T(\gvec{\rho}, t)$ first grows roughly as $t^{1/2}$ and then converges to a constant value $\sigma_{\infty}^2 \propto L \xi$, in sharp contrast to the relation $\sigma^2(t) \propto D t$ valid in the diffuse regime. Again, the localization length $\xi$ can be extracted directly from $\sigma^2(t)$, but contrary to the stationary case,  $\sigma^2(t)$ does not depend on absorption. Finally, exactly at the mobility edge, $\sigma^2(t)$ initially grows as $t^{1/2}$ as in the localized regime, until it saturates at $\sigma_{\infty}^2 \approx L^2$ in the long-time limit.

\begin{figure}
\includegraphics[width=12cm]{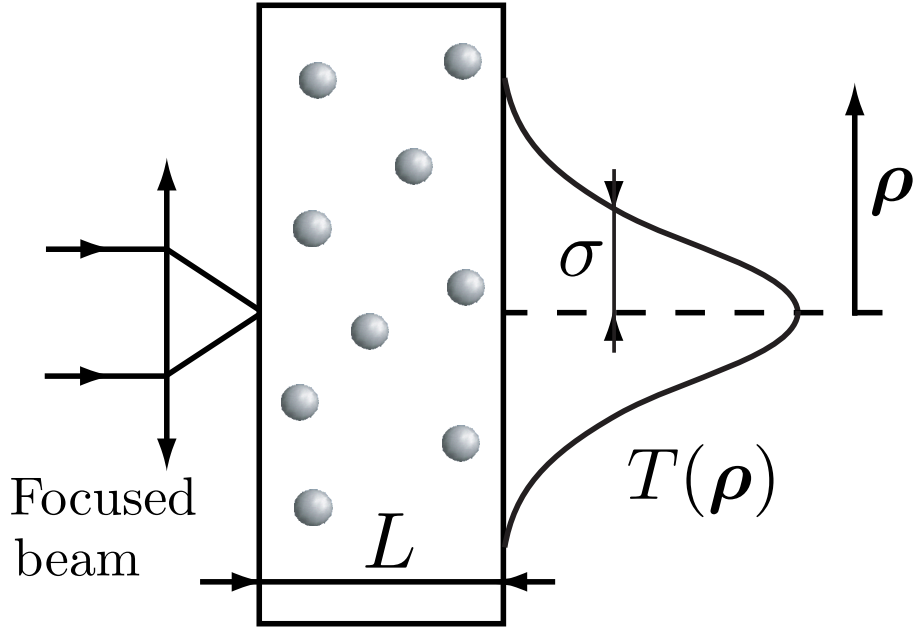}
\caption{\label{fig1} A beam is focused on the surface of 3D disordered slab of thickness $L$. The bell-shaped average position-dependent transmission coefficient $T(\gvec{\rho})$ has a mean square width $\sigma^2$ that depends on the strength of disorder in the medium.}
\end{figure}

A beam focused to a point $\gvec{\rho} = 0$ at the surface $z = 0$ of a disordered slab produces a bell-shaped intensity profile $T(\gvec{\rho})$ at the opposite face $z = L$ of the slab, as we schematically show in Fig.\ \ref{fig1}. We analyze $T(\gvec{\rho})$ in the framework on the self-consistent (SC) theory of localization \cite{vollhardt92}, extended to account for finite-size effects \cite{tiggelen00,skipetrov04,cherroret08}. The intensity Green's function $C(\vec{r}, \vec{r}^{\prime}, \Omega)$ obeys a diffusion equation
\begin{eqnarray}
\left[ -i \Omega - \nabla D(\vec{r}, \Omega) \nabla \right] C(\vec{r},
\vec{r}^{\prime}, \Omega) = \delta(\vec{r} - \vec{r}^{\prime})
\label{difeq}
\end{eqnarray}
with a self-consistently determined, position-dependent diffusion coefficient $D(\vec{r}, \Omega)$:
\begin{eqnarray}
\frac{1}{D(\vec{r}, \Omega)} = \frac{1}{D_B} +
\frac{12 \pi}{k^2 \ell} C(\vec{r}, \vec{r}, \Omega).
\label{d}
\end{eqnarray}
Here $k = 2 \pi/\lambda$ is the wavenumber, $\lambda$ is the wavelength, $\ell$ is the mean free path, and $D_B$ is the Boltzmann diffusion coefficient.
Equation (\ref{difeq}) applies to thick slabs with $L \gg \ell$ and should be solved with boundary conditions $C \mp  z_0 (D/D_B) \partial_z C = 0$ at the surfaces $z=0$ and $z=L$ \cite{cherroret08}. These boundary conditions are a generalization of the well-known boundary conditions for diffuse waves [$k\ell \gg 1$, $D(\vec{r}, \Omega)=D_B$] \cite{zhu91} to a medium with a position-dependent $D$. The length scale $z_0$ allows to account for internal reflections at the sample surface:
$z_0 = 2\ell/3$ in the absence of internal reflections and $z_0 > 2\ell/3$ when internal reflections are present. The exact value of $z_0$ can be found by specifying precise conditions at the sample boundary \cite{zhu91}. It remains of the order of $\ell$ under realistic experimental conditions.

The position-dependent diffusion coefficient $D(\vec{r}, \Omega)$ arises naturally from the microscopic derivation of SC theory in a bounded medium \cite{cherroret08}, in contrast to the previously employed concept of scale-dependent $D = D(\vec{q}, \Omega)$ (see, e.g., Refs.\ \cite{abrahams86} and \cite{chalker90}). In the slab geometry, $D(\vec{r}, \Omega) = D(z, \Omega)$ due to the translational invariance of the return probability $C(\vec{r}, \vec{r}, \Omega)$ (which is an ensemble-averaged quantity) in the $xy$ plane.
$T(\gvec{\rho})$ is found as $T(\gvec{\rho}) = -D(z=L,0) \partial_z C(\{\gvec{\rho}, z=L \}, \{\gvec{\rho}^{\prime} = 0, z^{\prime}=\ell \}, 0)$. We now focus on the mean square width of the transmitted beam: $\sigma^2 = \int \rho^2 T(\gvec{\rho}) d^2\gvec{\rho}/\int T(\gvec{\rho}) d^2\gvec{\rho}$ and consider three different regimes of wave propagation: weak disorder (diffusion, $k \ell \gg 1$), strong disorder (localization, $k \ell < 1$) and the critical regime (mobility edge, $k \ell = 1$).
To simplify the analysis, we assume $z_0 = 0$. We checked that non-zero values of $z_0$ only yield small corrections of the order of $z_0/L \sim \ell/L \ll 1$ to our results. This
might appear surprising because we know from the previous work that, at least in the diffuse regime [$k \ell \gg 1$, $D(\vec{r}, \Omega) = D_B$], $z_0$ may significantly affect certain measurable quantities. For example, the average stationary transmission coefficient of a disordered slab is
$T = \int T(\gvec{\rho}) d^2\gvec{\rho} = (\ell + z_0)/(L + 2 z_0)$. By neglecting $z_0$ one would make an error of about $66\%$ for the absolute value of $T$, even when $L \gg z_0$. One of the advantages of studying the mean square width $\sigma^2$ as defined above resides in its independence of the magnitude of $T(\gvec{\rho})$. $\sigma^2$ is only sensitive to the profile of $T(\gvec{\rho})$ as a function of $\gvec{\rho}$. This makes it virtually independent of $z_0$ as long as $z_0 \ll L$.

In the limit of weak disorder ($k \ell \gg 1$) the position dependence of $D$ can be neglected and one can set $D(z,0) = D_B [1 - (k\ell)^{-2}]$. Straightforward solution of the diffusion equation then yields
\begin{eqnarray}
T(\vec{q}) = \frac{\sinh q\ell}{\sinh qL}
\label{tqdif}
\end{eqnarray}
for the Fourier transform of $T(\gvec{\rho})$ and
\begin{eqnarray}
\sigma_{\mathrm{dif}}^2 &=&
\left. \frac{-\frac{1}{q} \frac{\partial}{\partial q}
\left[q \frac{\partial T(q)}{\partial q} \right]}{T(q)}
\right|_{q = 0}
=
\frac{2 L^2}{3} \left[ 1 - \left(\frac{\ell}{L} \right)^2
\right].
\label{dif}
\end{eqnarray}
For strong disorder ($k \ell < 1$), our self-consistent equations do not have an exact analytical solution. We set $D(z,0) \simeq D(0,0) \exp(-2\tilde{z}/\xi)$ with $\tilde{z} = \min(z, L-z)$, which is an ansatz inspired by the observation that this $D(z,0)$ represents an asymptotically exact solution in a semi-infinite medium for $z \gg \xi$ \cite{tiggelen00}. The expressions for $T(\vec{q})$ and $\sigma^2$ that follow from this ansatz are
\begin{eqnarray}
T(\vec{q}) &=& \frac{\xi \chi(q) e^{\ell/\xi} \sinh[\ell \chi(q)]}{\cosh[L \chi(q)] + \xi \chi(q) \sinh[L \chi(q)]-1},
\label{tqloc}
\\
\sigma_{\mathrm{loc}}^2 &=& 2 L \xi \left[
1 + \frac{1}{e^{L/\xi} - 1} -
\frac{\xi}{2 L} \left( 1 - e^{-L/\xi} \right)
- \frac{\ell}{L} \mathrm{Coth}\left(\frac{\ell}{\xi}\right) \right],
\label{loc}
\end{eqnarray}
where $\chi(q) = \sqrt{q^2 + 1/\xi^2}$.

At the mobility edge ($k \ell = 1$) an approximate expression for $D$ is $D(z,0) \simeq D(0,0)/(1 + \tilde{z}/z_c)$. Just like in the case of strong disorder, this approximation for $D$ is inspired by the solution for semi-infinite medium \cite{tiggelen00} and deviates from the numerical solution in the central part of the slab only. $D(0,0)$ and $z_c$ depend on the exact value of $z_0$ in the boundary conditions; $D(0,0) = D_B$ and $z_c \simeq 3 \ell$ for $z_0 = 0$ that we consider here. This yields
\begin{eqnarray}
T(\vec{q}) &=& \frac{\ell + z_c}{q z_c (L + 2 z_c)}
\frac{\eta_1(q)}{\eta_2(q) \eta_3(q)},
\label{tqme} \\
\sigma_{\mathrm{ME}}^2 &=&
\frac{3 L^2}{8} \left(1 + \frac{4 z_c}{L} \right),
\label{me}
\end{eqnarray}
where terms of order $(\ell/L)^2 \ll 1$ and $(z_c/L)^2 \ll 1$ were neglected in the brackets of Eq.\ (\ref{me}) and
\begin{eqnarray}
\eta_1(q) &=& I_1(q z_c) K_1[q(\ell + z_c)] -
K_1(q z_c) I_1[q(\ell + z_c)],
\label{eta1}
\\
\eta_2(q) &=& I_1(q z_c) K_0[q(L/2 + z_c)] +
K_1(q z_c) I_0[q(L/2 + z_c)],
\label{eta2}
\\
\eta_3(q) &=& I_1(q z_c) K_1[q(L/2 + z_c)] -
K_1(q z_c) I_1[q(L/2 + z_c)],
\label{eta3}
\end{eqnarray}
with $I_j$ and $K_j$ denoting modified Bessel functions of the first and second kinds, respectively.
We note that Eq.\ (\ref{me}) is not in contradiction with the multifractal statistics of wavefunctions at the mobility edge (see Ref.\ \cite{evers08} for a review) even though a link between $\sigma_{\mathrm{ME}}^2$ and multifractality is beyond the scope of this paper.

\begin{table}
\begin{tabular}{>{\centering}p{3.0cm}|>{\centering}p{3.0cm}|>{\centering}p{3.0cm}}
$k\ell$ & $D(z,0)$ & $\sigma^2$ \tabularnewline
\hline\hline
$k\ell\gg1$\\ (diffusion)&
$D_B \left[1 - \frac{1}{(k\ell)^{2}} \right]$&
$\dfrac{2}{3} L^2-\mathcal{O}(\ell^2)$
\tabularnewline
\hline
$k\ell=1$\\ (mobility edge)&
$\dfrac{D_B}{1 + \tilde{z}/z_c}$&
$\dfrac{3}{8}L^2+\mathcal{O}(L z_c)$
\tabularnewline
\hline
$k\ell<1$\\ (localization)&
$D_B e^{-2\tilde{z}/\xi}$&
$2 L \xi + \mathcal{O}(\xi^2)$
\tabularnewline
\end{tabular}
\caption{\label{table1} Summary of analytical results for the position-dependent diffusion coefficient $D(z,0)$ and its associated mean square width $\sigma^2$ of the position-resolved transmission coefficient $T(\gvec{\rho})$ describing transmission of a tightly focused monochromatic beam through a disordered slab of thickness $L \gg \ell$. $\ell$ is the mean free path, $k$ is the wave number, $z_c \simeq 3 \ell$, $\xi$ is the localization length, and $\tilde{z} = \min(z, L-z)$. We assume $z_c \ll L$ for $k \ell = 1$ and $\ell \ll \xi \ll L$ for $k \ell < 1$.}
\end{table}

\begin{figure}
\includegraphics[width=12cm]{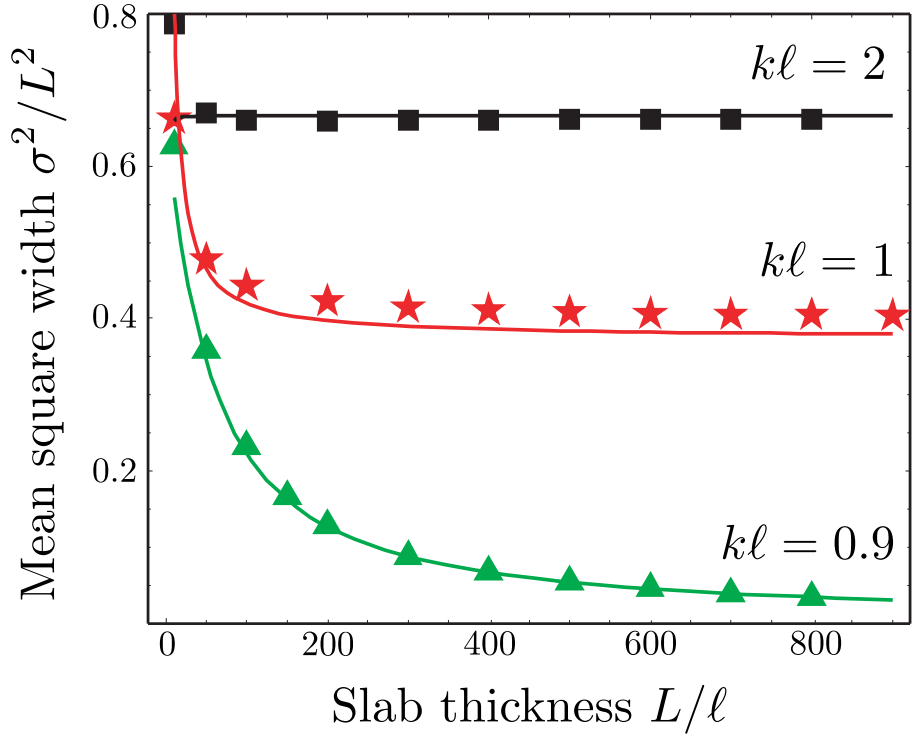}
\caption{\label{fig2}
Comparison of approximate analytical (lines) and exact numerical (symbols) results for the mean square width $\sigma^2$ of the spatially resolved stationary transmission coefficient $T(\gvec{\rho})$ of a slab.}
\end{figure}

We summarize the analytical results for $\sigma^2$ following from Eqs.\ (\ref{dif}), (\ref{loc}) and (\ref{me}) in the limit of large $L$ in Table \ref{table1}. Because these results are derived using approximate expressions for $D(z,0)$, a comparison with exact numerical solutions is required to justify their validity. To this end, we solve the SC equations of localization numerically and show the resulting mean square widths $\sigma^2$ as functions of slab thickness $L$ in Fig.\ \ref{fig2}. The agreement between analytical and numerical results is satisfactory, confirming the validity of our analytical analysis.

An important comment is in order. It is remarkable that in the case of weak disorder ($k \ell \gg 1$), $\sigma^2 = 2 L^2/3$ does \emph{not} depend on $D$ and even a very small, but finite and spatially uniform diffusion coefficient would lead to the same result for $\sigma^2$. This emphasizes the importance of the position dependence of $D(z,0)$ in the localized regime $kl < 1$ and at the mobility edge $kl = 1$. Note that this is the first time that the position dependence of $D$ is absolutely \emph{vital} to obtain a result that is different from the universal diffuse outcome. Scale-dependent but spatially uniform diffusion coefficients $D \propto 1/L$ and $D \propto \exp(-L/\xi)$ put forward by the scaling theory \cite{abrahams79} explain the scaling of the stationary transmission coefficient $T \propto 1/L^2$ at the mobility edge and $T \propto \exp(-L/\xi)$ in the localized regime, respectively. However, being uniform in space, these expressions for $D$ result in exactly the same mean square width of the transmitted beam as $D = D_B$: $\sigma^2 = 2 L^2/3$.
It is the position dependence of $D$ that accounts for the new expressions for $\sigma^2$ that we find at strong disorder. Although $\sigma^2 = 3 L^2/8$ at the mobility edge differs from the result in the diffuse regime by just a numerical coefficient, in the localized regime $\sigma^2 \simeq 2 L \xi$ links $\sigma$ to the localization length $\xi$, which offers an elegant way of measuring the latter experimentally.

Absorption was a serious obstacle for the unambiguous interpretation of a number of experiments on Anderson localization \cite{wiersma97,scheffold99,schuur99,chabanov00}. It is therefore important to study the role of absorption in the context of transverse confinement of waves in 3D. A straightforward calculation in the regime of weak disorder ($k \ell \gg 1$) shows that if the macroscopic absorption length $L_a = \sqrt{\ell/3\mu_a}$ (where $\mu_a$ is the absorption coefficient) is much longer than the sample thickness $L$, $\sigma^2$ given in Table\ \ref{table1} only acquires a small correction $-2 L^4/45 L_a^2$. Hence, weak absorption is not an obstacle for observing transverse confinement. However, in the opposite case of $L_a \ll L$ we obtain $\sigma^2 \simeq 2 L L_a$, i.e. absorption plays \emph{exactly} the same role as localization and the resulting equation for $\sigma^2$ coincides with the one obtained in the localized regime, with $\xi$ replaced by $L_a$. This indicates that a study of stationary transverse confinement cannot distinguish localization from absorption and hence suffers from the same drawbacks as previous works \cite{wiersma97,scheffold99,schuur99,chabanov00}.

A way to overcome complications due to absorption is suggested by recent works \cite{chabanov03,storzer06,zhang07,skipetrov04}. The idea is to study the \textit{dynamics} of wave propagation rather than the stationary transport. We adopt this idea here too and replace the continuous incident beam in the experiment depicted in Fig.\ \ref{fig1} by a short pulse.
We assume that the duration of the pulse $t_p$ is, on the one hand, much shorter than the typical time
$t_D = (L + 2 z_0)^2/\pi^2 D_B$ required for the wave to cross the disordered sample, but, on the other hand, not too short to ensure that the frequency-dependent properties of the disordered medium (such as the mean free path $\ell$, the localization length $\xi$ or the absorption coefficient $\mu_a$) do not change significantly within the frequency band $1/t_p$ of the pulse. This assumption allows us to treat the incident pulse as a delta-pulse $\delta(t)$ in Eq.\ (\ref{difeq}) without introducing additional complications due to sample-specific dependencies of $\ell$, $\xi$ and $\mu_a$ on the frequency within the bandwidth of the pulse. Corresponding experimental conditions were realized, for example, in Refs.\ \cite{zhang07} and \cite{cherroret09b} with microwaves and Refs.\ \cite{hu08} and \cite{page95} with ultrasound.

\begin{figure}
\includegraphics[width=12cm]{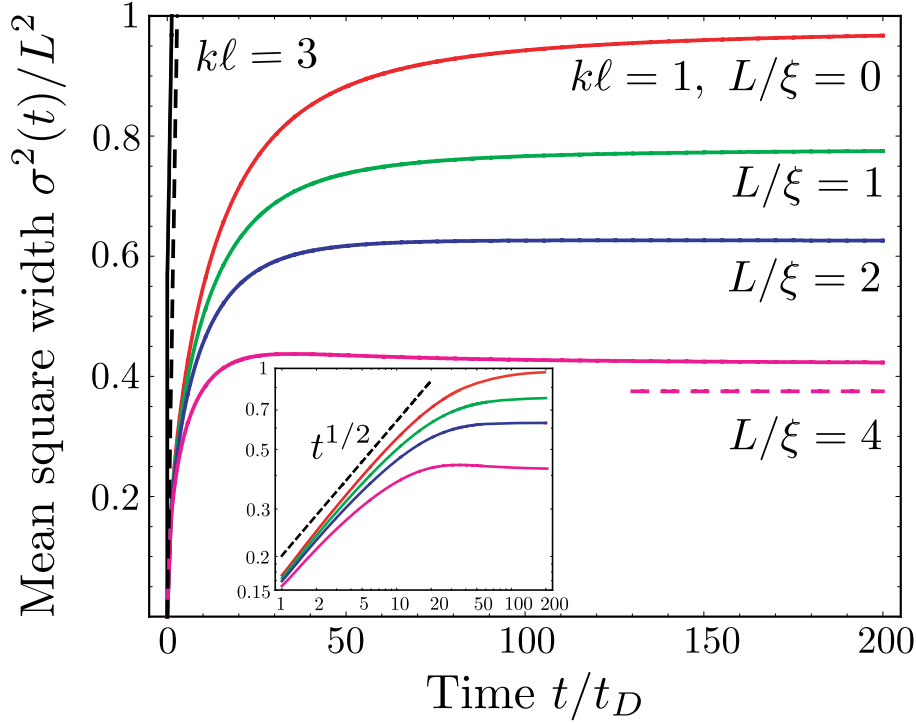}
\caption{\label{fig3}
Numerically calculated mean square width $\sigma^2(t)$ of the time-dependent transmission coefficient $T(\gvec{\rho}, t)$ of a disordered slab in the diffuse regime ($k \ell = 3$), at the mobility edge ($k \ell = 1$), and in the localized regime ($L/\xi = 1$, 2, 4). The dashed lines show $\sigma^2(t) = 4 D_B [1 - (k\ell)^{-2}] t$ for $k \ell = 3$ and $\sigma_{\infty}^2 = 2 L \xi (1 - \xi/L)$ for $L/\xi = 4$. The thickness of the slab is $L = 100\ell$. Time is given in units of $t_D = (L + 2 z_0)^2/\pi^2 D_B$. The inset shows the same results in the log-log scale, the dashed line is $\sigma^2(t) \propto t^{1/2}$.}
\end{figure}

Considering a short pulse instead of a continuous wave complicates the analysis considerably because no simple analytic approximation exists for $D(z, \Omega)$ at arbitrary $\Omega \ne 0$, contrary to the stationary $\Omega = 0$ case. One way to calculate $\sigma^2$, which now acquires a time dependence, is to solve the SC equations of localization numerically. Because simplified boundary conditions with $z_0 = 0$ used previously do not introduce any simplification in the numerical analysis, we restore the finite value of $z_0$ here and set $z_0 = 2\ell/3$. The results of the numerical solution of Eqs.\ (\ref{difeq}) and (\ref{d}) are shown in Fig.\ \ref{fig3}.
The difference between diffuse and localized regimes is manifest in Fig.\ \ref{fig3}. The rise of $\sigma^2(t)$ with time is unbounded for $k \ell > 1$, whereas $\sigma^2(t)$ saturates at a finite value $\sigma_\infty^2$ for $k \ell < 1$. The latter result can be understood from the following approximate calculation which, however, turns out to be quite adequate. Instead of using the self-consistent equation for $D(\vec{r}, \Omega)$, let us simply set $D(z, \Omega) = -i \Omega \xi^2$, which is the solution of SC model in the infinite medium in the limit $\Omega \rightarrow 0$. A straightforward calculation then yields $\sigma_{\infty}^2 = 2 L \xi (1 - \xi/L)$ up to the first order in $\xi/L \ll 1$. This equation, shown by a dashed horizontal line in Fig.\ \ref{fig3}, falls fairly close to the numerical result at $L/\xi = 4$, allowing us to conjecture that it might be a good estimate of  $\sigma_{\infty}^2$ in the limit of $L \gg \xi$.
Our calculation suggests that the saturation of $\sigma^2(t)$ at a constant level takes place not only in the localized regime, but even at the mobility edge. We find the asymptotic value of $\sigma(t)$ to be $\sigma_{\infty} \approx L$ (see Fig.\ \ref{fig3}).

It is worthwhile to note that our time-dependent  $\sigma^2(t)$ is not sensitive to absorption and that all curves of Fig.\ \ref{fig3} remain exactly the same in an absorbing medium. Absorption cannot lead to bending and saturation of $\sigma^2(t)$ with time. In a medium with absorption but without localization effects the curves $\sigma^2(t)$ just grow linearly: $\sigma^2(t) = 4 D_B t$. The confusion between localization and absorption is therefore impossible if $\sigma^2(t)$ is studied. This provides a solution to the long-standing issue of distinguishing localization from absorption \cite{wiersma97,scheffold99,schuur99,chabanov00}, as has been recently demonstrated in Ref.\ \cite{hu08}.

Let us now analyze the time dependence of $\sigma^2(t)$. First, in the diffuse regime $k \ell \gg 1$ one readily obtains $\sigma^2(t) \simeq 4 D t$, where $D \simeq D_B[1 - (k\ell)^{-2}]$ can be assumed position-independent. For $k \ell = 3$ this result is shown in Fig.\ \ref{fig3} by a dashed line that is indeed very close to the result of the numerical calculation. To study the time dependence of $\sigma^2(t)$ at the mobility edge and in the localized regime, we replot the curves of the main plot, except the one corresponding to $k \ell = 3$,  in the inset of Fig.\ \ref{fig3} in a log-log scale. We clearly see that the initial growth of $\sigma^2(t)$ with $t$ is power-law: $\sigma^2(t) \propto t^{\alpha}$ with $\alpha \simeq 0.5$. It is remarkable that this power-law growth is observed not only in the localized regime but at the mobility edge $k \ell = 1$ as well. To understand the power exponent $\alpha \simeq 0.5$, let us consider a point source of waves located inside a random medium of size $L \gg \ell$, very far from its boundaries.
Instead of introducing the position dependence of $D$, one can use scaling arguments and calculate the return probability with a lower cutoff $q_{\mathrm{min}} \sim 1/L$ in the integration over the momentum \cite{vollhardt92,shapiro82}:
\begin{eqnarray}
C(\vec{r}, \vec{r}, \Omega) = \frac{1}{2 \pi^2}
\int_{q_{\mathrm{min}}}^{q_{\mathrm{max}}}
\frac{dq\; q^2}{-i \Omega + D(\Omega) q^2},
\label{return}
\end{eqnarray}
where $q_{\mathrm{max}}\sim 1/\ell$ ensures convergence of the integration. Inserting this expression into Eq.\ (\ref{d}) and solving for $D(\Omega)$ at $k \ell = 1$, we find $D(\Omega) \propto (-i \Omega)^{1/3}$ for $\Omega \gg D_B/L^2$ and $D(\Omega) \propto (-i \Omega)^{1/2}$ for $\Omega \ll D_B/L^2$. The typical mean square radius $\langle \vec{r}^2 \rangle$ of the intensity profile in 3D is then
$\langle \vec{r}^2 \rangle \propto t^{2/3}$ for $t \ll L^2/D_B$ and
$\langle \vec{r}^2 \rangle \propto t^{1/2}$ for $t \gg L^2/D_B$.
If we now assume that $\langle \vec{r}^2 \rangle$ and $\sigma^2(t) = \langle \gvec{\rho}^2 \rangle$ in transmission through a slab behave in a similar way, we obtain a qualitative explanation for the power-law scaling $\sigma^2(t) \propto t^{1/2}$ that we observe in Fig.\ \ref{fig3} at $t > t_D$.
Note, however, that this simple scaling argument does not explain the saturation of $\sigma^2(t)$ at longer $t$.

Another interesting and unexpected feature of $\sigma^2(t)$ that we found sufficiently deep in the localized regime is its nonmonotonity with time: the curve corresponding to $L/\xi = 4$ in Fig.\ \ref{fig3} has a weakly pronounced but clearly visible maximum at $t/t_D \simeq 30$. Even though this nonmonotonity is predicted by the SC model (we checked that the maximum is present at least for $L/\ell = 50$--$200$ and $L/\xi = 3$--$6$), it remains to be seen whether it is an artifact of the model or a real physical phenomenon.

Equations (\ref{difeq}) and (\ref{d}) that we used in this paper were derived under the assumption of white-noise, uncorrelated disorder \cite{cherroret08}. Following well-understood approaches \cite{akkermans07}, they can be generalized to the case of disorder with finite correlation length by replacing $\ell$ by the bare transport mean free path $\ell_B^{*}$, associated with diffusion in the absence of macroscopic interferences. In the context of SC theory of localization with a position-independent $D$, such a generalization was acomplished by W\"{o}lfle and Bhatt \cite{wolfle84}; see also Ref.\ \cite{vollhardt92}. In an open medium of finite size (slab) considered in the present paper, $\ell_B^{*}$ should also be substituted for $\ell$ in the boundary conditions for $C(\vec{r}, \vec{r}^{\prime}, \Omega)$ and in the expression $z^{\prime} = \ell$ for the position of the delta-source $\delta(z-z^{\prime})$ that models the source of diffuse radiation due to the incident wave. These two replacement are fully justified for diffuse waves \cite{zhu91} and are known to yield excellent agreement with stationary \cite{zhu91,li93} and dynamic \cite{page95,johnson03} transmission experiments, as well as with numerical simulations \cite{durian94}. Because in our SC model localization effects manifest themselves only through a reduced and position-dependent diffusion coefficient which, in addition, is assumed to vary slowly in space [i.e., $D(\vec{r}, \Omega)$ does not vary significantly on the scale of $\ell$ or $\ell_B^{*}$], the replacement $\ell \rightarrow \ell_B^{*}$ makes our results valid for media with correlated disorder, provided that $L \gg \ell_B^{*}$.
In the localized regime, the requirement of slow variation of $D(\vec{r}, \Omega)$ in space reduces to the condition $\xi \gg \ell, \ell_B^*$. This means that SC theory does not allow us to explore very strong localization, when $\xi$ becomes of the order of the mean free path.

It is important to stress that the SC theory does not yield the position of the mobility edge but rather requires it as an input. In the present paper we have chosen to put the mobility edge at $k \ell = 1$, but it is unclear how (and if) it should be shifted in the case of correlated disorder. When the results presented in this paper are generalized to media with correlated disorder by replacing $\ell$ by $\ell_B^{*}$, they correspond to the mobility edge located at $k \ell_B^{*} = 1$. The shift of the mobility edge might slightly affect certain numerical coefficients in our results and these coefficients should therefore be considered as approximate. This limitation arises from the uncertainty in the exact location of the mobility edge intrinsic to the SC theory and it is common for the cases of both white-noise and correlated disorders.

Another known issue of SC theory is the discrepancy of the critical exponent $\nu = 1$ that it yields with the results of numerical solution of Anderson tight-binding model: $\nu \approx 1.5$ \cite{num}. On the one hand, this may suggest that the use of SC theory is questionable in the vicinity of the mobility edge. But on the other hand, the SC theory correctly reproduces some other features of critical behavior (like, e.g., the anomalous diffusion $\langle \vec{r}^2 \rangle \propto t^{2/3}$ \cite{shapiro82} or the $1/L^2$ scaling of the transmission coefficient $T$ of a slab \cite{vollhardt82}) which seem to be more relevant in the present context. This demonstrates that the SC theory is capable of producing correct results for both stationary and dynamic properties of wave transport even at the mobility edge. Recently, a modification of SC theory has been proposed that allows to obtain $\nu$ close to the numerical value at a price of introducing scale ($q$) dependence of the diffusion coefficient $D$, based on phenomenological considerations \cite{garcia08}. This approach is different from our concept of position-dependent $D$ that has an advantage of being justified microscopically \cite{cherroret08}. It would be interesting to compare predictions of the approach of Ref.\ \cite{garcia08} for the transverse confinement problem with our results. And of course, direct numerical simulations of either the Anderson model \cite{num,am} or the classical wave equation \cite{we} might be a crucial test for our conclusions. Even though simulations of large 3D disordered systems are time- and memory-consuming, the constant growth of available computer power makes them more and more accessible.

In conclusion, we have shown that the transverse confinement of waves in 3D random media can be very useful for demonstrating Anderson localization and for measuring the localization length in an experiment. The dependence of the mean square width $\sigma^2$ of a tightly focused beam transmitted through a slab of random medium on the slab thickness $L$ and on time (in a pulsed experiment) is qualitatively different in the diffuse and in the localized regimes of wave propagation. In a pulsed experiment, $\sigma^2(t)$ is independent of absorption in the disordered sample and thus measuring $\sigma^2(t)$ allows to avoid the risk of confusing localization and absorption.

We thank J.H. Page for stimulating discussions.
The computations presented in this paper were performed on the cluster PHYNUM (CIMENT, Grenoble). S.E.S. acknowledges financial support from the French ANR (project No. 06-BLAN-0096 CAROL) and the French Ministry of Education and Research.


\begin{thebibliography}{99}

\bibitem{anderson58}
P.W. Anderson,
\textit{Phys. Rev.} \textbf{109}, 1492 (1958).

\bibitem{john84}
S. John,
\textit{Phys. Rev. Lett.} \textbf{53}, 2169 (1984).

\bibitem{anderson85}
P.W. Anderson,
\textit{Phil. Mag. B} \textbf{52}, 505 (1985).

\bibitem{bart99}
B.A. van Tiggelen,
in \textit{Diffuse Waves in Complex Media}, edited by J. P. Fouque (Kluwer, Dordrecht, 1999), p. 1.

\bibitem{wiersma97}
D.S. Wiersma \textit{et al.},
\textit{Nature} \textbf{390}, 671 (1997).

\bibitem{scheffold99}
F. Scheffold \emph{et al.},
\textit{Nature} \textbf{398}, 206 (1999);
D.S. Wiersma \emph{et al.},
\textit{Nature} \textbf{398}, 207 (1999).

\bibitem{schuur99}
F.J.P. Schuurmans, M. Megens, D. Vanmaekelbergh, and A. Lagendijk,
\textit{Phys. Rev. Lett.} \textbf{83}, 2183 (1999).

\bibitem{chabanov00}
A.A. Chabanov, M. Stoytchev, and A.Z. Genack,
\textit{Nature} \textbf{404}, 850 (2000).

\bibitem{chabanov03}
A.A. Chabanov, Z.Q. Zhang, and A.Z. Genack,
\textit{Phys. Rev. Lett.} \textbf{90}, 203903 (2003).

\bibitem{skipetrov04}
S.E. Skipetrov and B.A. van Tiggelen,
\textit{Phys. Rev. Lett.} \textbf{92}, 113901 (2004);
\textit{ibid.} \textbf{96}, 043902 (2006).

\bibitem{storzer06}
M. St\"{o}rzer, P. Gross, C.M. Aegerter, and G. Maret,
\textit{Phys. Rev. Lett.} \textbf{96}, 063904 (2006).

\bibitem{zhang07}
Z.Q. Zhang, A.A. Chabanov, S.K. Cheung, C.H. Wong, and A.Z. Genack,
\textit{Phys. Rev. B} \textbf{79}, 144203 (2009).

\bibitem{schwarz07}
T. Schwarz, G. Bartal, Sh. Fishman, and M. Segev,
\textit{Nature} \textbf{456}, 52 (2007).

\bibitem{raedt89}
H. de Raedt, A. Lagendijk, and P. de Vries,
\textit{Phys. Rev. Lett.} \textbf{62}, 47 (1989).

\bibitem{abrahams79}
E. Abrahams, P.W. Anderson, D.C. Licciardello, and T.V. Ramakrishnan,
\textit{Phys. Rev. Lett.} \textbf{42}, 673 (1979).

\bibitem{hu08}
H. Hu, A. Strybulevych, J.H. Page, S.E. Skipetrov, and B.A. van Tiggelen,
\textit{Nature Physics} \textbf{4}, 945 (2008).

\bibitem{tiggelen00}
B.A. van Tiggelen, A. Lagendijk, and D.S. Wiersma,
\textit{Phys. Rev. Lett.} \textbf{84}, 4333 (2000).

\bibitem{cherroret08}
N. Cherroret and S.E. Skipetrov,
\textit{Phys. Rev. E} \textbf{77}, 046608 (2008).

\bibitem{vollhardt92}
D. Vollhardt and P. W\"{o}lfle,
in \textit{Electronic Phase Transitions} (Elsevier Science, Amsterdam, 1992), p.1.

\bibitem{zhu91}
J.X. Zhu, D.J. Pine, and D.A. Weitz,
\textit{Phys. Rev. A} \textbf{44}, 3948 (1991);
R.C. Haskell, L.V. Swaasand, T. Tsay, T. Feng, M.S. McAdams,
and B.J. Tromberg,
\textit{J. Opt. Soc. Am. A} \textbf{11}, 2727 (1994).

\bibitem{abrahams86}
E. Abrahams and P.A. Lee,
\textit{Phys. Rev. B} \textbf{33}, 683 (1986).

\bibitem{chalker90}
J.T. Chalker,
\textit{Physica A} \textbf{167}, 253 (1990).

\bibitem{evers08}
F. Evers and A.D. Mirlin,
\textit{Rev. Mod. Phys.} \textbf{80}, 1355 (2008).

\bibitem{li93}
J.H. Li, A.A. Lisyansky, T.D. Cheung, D. Livdan and A.Z. Genack,
\textit{Europhys. Lett.} \textbf{22}, 675 (1993).

\bibitem{cherroret09b}
N. Cherroret, A. Pe{\~{n}}a, A.A. Chabanov and S.E. Skipetrov, \textit{Phys. Rev. B} \textbf{80}, 045118 (2009).

\bibitem{page95}
J.H. Page, H.P. Schriemer, A.E. Bailey, and D.A. Weitz,
\textit{Phys. Rev. E} \textbf{52}, 3106 (1995).

\bibitem{shapiro82}
B. Shapiro,
\textit{Phys. Rev. B} \textbf{25}, 4266 (1982).

\bibitem{akkermans07}
E. Akkermans and G. Montambaux,
\emph{Mesoscopic Physics of Electrons and Photons}
(Cambridge University Press, Cambridge, 2007).

\bibitem{wolfle84}
P. W\"{o}lfle and R.N. Bhatt,
\textit{Phys. Rev. B} \textbf{30}, 3542 (1984).

\bibitem{johnson03}
P.M. Johnson, A. Imhof, B.P. Bret, J. G\'{o}mez Rivas, and A. Lagendijk,
\textit{Phys. Rev. E} \textbf{68}, 016604 (2003).

\bibitem{durian94}
D.J. Durian,
\textit{Phys. Rev. E} \textbf{50}, 857 (1994)

\bibitem{num}
M. Schreiber and H. Grussbach,
\textit{Phys. Rev. Lett.} \textbf{76}, 1687 (1996);
K. Slevin and T. Ohtsuki,
\textit{Phys. Rev. Lett.} \textbf{82}, 382 (1999).

\bibitem{vollhardt82}
D. Vollhardt and P. W\"{o}lfle,
\textit{Phys. Rev. Lett.} \textbf{48}, 699 (1982);
N. Cherroret, S.E. Skipetrov and B.A. van Tiggelen,
\textit{Phys. Rev. B} \textbf{80}, 037101 (2009).

\bibitem{garcia08}
A.M. Garc\'{i}a-Garc\'{i}a,
\textit{Phys. Rev. Lett.} \textbf{100}, 076404 (2008).

\bibitem{am}
M. Weiss, J.A. M\'{e}ndez-Berm\'{u}dez, and T. Kottos,
\textit{Phys. Rev. B} \textbf{73}, 045103 (2006);
J. Brndiar and P. Markos,
\textit{Phys. Rev. B} \textbf{77}, 115131 (2008);
A. Rodriguez, L.J. Vasquez, and R.A. R\"{o}mer,
\textit{Phys. Rev. Lett.} \textbf{102}, 106406 (2009).

\bibitem{we}
C. Conti, L. Angelani, and G. Ruocco,
\textit{Phys. Rev. A} \textbf{75}, 033812 (2007);
C. Conti, A. Fratalocchi,
\textit{Nature Physics} \textbf{4}, 794 (2008);
A. Sheikhan, M. Reza Rahimi Tabar, and M. Sahimi,
\textit{Phys. Rev. B} \textbf{80}, 035130 (2009).

\end{thebibliography}
\end{document}